\documentclass[11pt,a4paper]{article}

\usepackage{amsmath,amssymb,tikz}
\usepackage{graphicx}
\usepackage{amsmath}
\usepackage{wrapfig}
\usepackage{float}
\usepackage{verbatim}
\usepackage{titlesec}
 \allowdisplaybreaks
\long\def\symbolfootnote[#1]#2{\begingroup%
\def\thefootnote{\fnsymbol{footnote}}\footnote[#1]{#2}\endgroup}

\newcommand{\sysu}{{\it School of Physics and Astronomy, Sun Yat-Sen University, 2 Daxue Road, Zhuhai 519082, China}}

\begin{document}
\thispagestyle{empty}
\begin{center}

~\vspace{20pt}

{\Large\bf Traversable Wormhole in AdS and Entanglement}

\vspace{25pt}

	{\large Hua-Chao Liu\footnote{Email: \texttt{liuhch39@mail2.sysu.edu.cn}},
	\large Rong-Xin Miao\footnote{Email: \texttt{miaorx@mail.sysu.edu.cn}}}

\vspace{10pt}${}$\sysu

\vspace{2cm}

\begin{abstract}
A traversable wormhole generally violates the averaged null energy condition, usually requiring exotic matter. Recently, it has been found that the traversable wormhole can be realized by non-exotic matter in Einstein-Dirac-Maxwell theories in flat space. This paper generalizes discussions to the AdS spacetime and finds traversable wormholes with spherical and planar topologies. Furthermore, based on the AdS/CFT correspondence, we compute the entanglement entropy of strips and disks on two AdS boundaries of the wormhole. We find that entanglement entropy undergoes a phase transition as the subsystem size increases. 
\end{abstract}

\end{center}

\newpage
\setcounter{footnote}{0}
\setcounter{page}{1}

\tableofcontents
\newpage
\section{Introduction}

Like a black hole, a wormhole is a fascinating solution to Einstein's equations that has garnered significant interest in the theoretical physics community. Notably, it plays a crucial role in the recent breakthroughs regarding the black hole information paradox \cite{Penington:2019kki,Almheiri:2019qdq}. Generally, there are two types of wormholes: nontraversable and traversable. A well-known example of a nontraversable wormhole is the Einstein-Rosen bridge \cite{Einstein:1935tc}, which connects two asymptotic regions of eternal black holes. According to the ``ER=EPR" conjecture \cite{Maldacena:2013xja}, it is dual to entangled Einstein–Podolsky–Rosen (EPR) pairs, indicating a profound connection between wormholes and quantum entanglement. The traversable wormhole was first studied by Ellis \cite{Ellis:1973yv} and Bronnikov \cite{Bronnikov:1973fh} and then by Morris and Thorne \cite{Morris:1988cz}. However, these traversable wormholes generally violate the averaged null energy condition (ANEC), which raises questions about their existence. It's important to note that the violation of ANEC can be accommodated through quantum fluctuations, such as the Casimir effect, which is considered acceptable. Nevertheless, the traversable wormhole described by Morris and Thorne allows for time travel to the past, which contradicts the law of causality \cite{Morris:1988tu}. The first traversable wormhole consistent with causality was proposed by Gao, Jafferis, and Wall using a double trace deformation \cite{Gao:2016bin}. In this model, it takes longer to travel through the wormhole throat than it does via the typical path outside. Thus, this traversable wormhole adheres to the principles of causality and may be realized in the physical world. A significant advancement in this area is the humanly traversable wormholes proposed by Maldacena, Milekhin, and Popov, which utilize Casimir-like energy from fermions or the Randall-Sundrum brane-world scenario \cite{ Maldacena:2018gjk,Maldacena:2020sxe}. Following this, wormhole solutions to Einstein's gravity coupled with Maxwell fields and two Dirac fermions were discovered \cite{Blazquez-Salcedo:2020czn} and refined in subsequent studies \cite{Konoplya:2021hsm, Wang:2022aze}. Remarkably, this traversable wormhole is composed entirely of non-exotic matter. See also \cite{Kundu:2021nwp,DeFalco:2023twb,Battista:2024gud,Bouhmadi-Lopez:2021zwt} for some recent developments of wormholes. 

This paper extends the findings of \cite{Blazquez-Salcedo:2020czn, Konoplya:2021hsm} from flat space to AdS space. Subsequently, we are able to explore the CFT duals of traversable wormholes within the framework of the AdS/CFT correspondence \cite{Maldacena:1997re}. We numerically solve the AdS wormholes with spherical and planar topologies in Einstein-Dirac-Maxwell theories. We confirm that these solutions violate the NEC and are, therefore, genuine wormhole solutions. Additionally, we examine the holographic entanglement entropy (HEE) \cite{Ryu:2006bv} of strips and disks on the two AdS boundaries of the wormholes. As the size of the strip or disk grows, the Ryu-Takayanagi (RT) surface \cite{Ryu:2006bv} for entanglement entropy undergoes a phase transition, transitioning from a disconnected state to a connected one. Intriguingly, for the disk, the connected extremal surface only emerges when the disk radius surpasses a critical value. We validate this peculiar phenomenon using a toy model, verifying its occurrence in the case of the planar wormhole as well.

Let us state the motivations for investigating the HEE in the wormhole background. First, HEE plays an essential role in AdS/CFT. It reveals a deep relation between the geometry and the quantum entanglement \cite{Ryu:2006bv}. From the first law of entanglement entropy, one can derive the linear Einstein equations in the framework of AdS/CFT \cite{Lashkari:2013koa,Faulkner:2013ica}. Furthermore, the recent breakthrough in the black hole information problem also benefits from the developments of HEE, particularly the concept of entanglement islands \cite{Penington:2019kki,Almheiri:2019qdq}. Second, ``ER=EPR" shows that the nontraversable wormhole is closely related to the entanglement entropy \cite{Maldacena:2013xja}. Thus, it is interesting to study if there are also some relations between the traversable wormhole of this paper and the entanglement entropy. Third, the fermions on the two sides of the flat wormhole are entangled \cite{Blazquez-Salcedo:2020czn}. Thus, exploring the entanglement of two sides of the AdS wormhole is interesting, as we have done in this paper. 

This paper is organized as follows. In section 2, we briefly review the Einstein-Dirac-Maxwell (EDM) model. In section 3, we numerically solve the EDM model to obtain traversable AdS wormholes with spherical and planar topologies, respectively. Section 4 studies the HEE of strips in the AdS wormhole and discusses the phase transition. Section 5 generalizes the discussions to HEE of disks. Finally, we conclude with some discussions in section 6. 

\label{sec:intro}



\section{Einstein-Dirac-Maxwell model}

This section gives a quick review of the Einstein-Dirac-Maxwell model, which contains one vector field $A_{\mu}$ and two spinor fields $\Psi_1$ and $\Psi_2$. The action reads \cite{Blazquez-Salcedo:2020czn}
\begin{align}\label{action}
	I=\frac{1}{4\pi}\int d^4x\sqrt{-g}\left[\frac{1}{4}(R+\frac{6}{l^2})+\mathcal{L}_{D}-\frac{1}{4}\mathcal{F}^2\right],
\end{align}
where we have set Newton's constant $G_N=1$, $R$ is the Ricci scalar, $l$ is the AdS radius, $\mathcal{F}_{\mu\nu}=\partial_{\mu}A_{\nu}-\partial_{\nu}A_{\mu}$ is the field strength of the vector, and
\begin{align}\label{matterA}
	\mathcal{L}_D=\sum_{\epsilon=1,2}\left[\frac{i}{2}\bar{\Psi}_{\epsilon}\gamma^{\nu}\hat{D}_{\nu}\Psi_{\epsilon}-\frac{i}{2}\hat{D}_{\nu}\bar{\Psi}_{\epsilon}\gamma^{\nu}\Psi_{\epsilon}-m\bar{\Psi}_{\epsilon}\Psi_{\epsilon}\right].
\end{align}
Here $\gamma_{\mu}\equiv e_{\mu a}\hat{\gamma}^{a}$ and $\hat{\gamma}^{a}$ are the gamma matrices in curved and flat space respectively, $e_{\mu a}$ are tetrad fields, $m$ is the spinor mass, and 
\begin{align}            
	\label{DandGamma}
	\hat{D}_{\mu}&=\partial_{\mu}+\Gamma_{\mu}-iq A_{\mu},\\
	\Gamma_{\mu}&=-\frac{1}{4}\omega_{\mu a b}\hat{\gamma}^{a}\hat{\gamma}^{b},\end{align}
with $\omega_{\mu ab}$ the spin connections. The gamma matrices in flat space read  \cite{Konoplya:2021hsm}
\begin{align}\label{gammaM}
	\hat{\gamma}^0=i\begin{pmatrix}
		0&I\\
		I&0
	\end{pmatrix},
	\hat{\gamma}^1=i\begin{pmatrix}
		0&\sigma_3\\
		-\sigma_3&0
	\end{pmatrix},
	\hat{\gamma}^2=i\begin{pmatrix}
		0&\sigma_2\\
		-\sigma_1&0
	\end{pmatrix},
	\hat{\gamma}^3=i\begin{pmatrix}
		0&\sigma_2\\
		-\sigma_2&0
	\end{pmatrix},\tag{7a}
\end{align}
where $\sigma_i$ denote the Pauli matrices
\begin{align}\label{pauliM}
	\sigma_1=\begin{pmatrix}
		0&1\\
		1&0
	\end{pmatrix},\quad
	\sigma_2=\begin{pmatrix}
		0&-i\\
		i&0
	\end{pmatrix},\quad
	\sigma_3=\begin{pmatrix}
		1&0\\
		0&-1
	\end{pmatrix}.\tag{7b}
\end{align}
\setcounter{equation}{7}
From the action (\ref{action}), we derive the equations of motion (EOM)
\begin{align} \label{EOM1}
	&\left(\gamma^{\mu}\hat{D}_{\mu}-m\right)\Psi_{\epsilon}=0,\tag{8a}\\
	\label{EOM2}
	&\frac{\partial_{\nu}\sqrt{-g} \mathcal{F}^{\mu\nu}}{2\sqrt{-g}}-qj^{\mu}=0\text{ with } j^{\mu}=\sum_{\epsilon=1,2}\bar{\Psi}_{\epsilon}\gamma^{\mu}\Psi_{\epsilon},\tag{8b} \\
	\label{EOM3}
	&\frac{1}{2}\left(R_{\mu\nu}-\frac{1}{2}g_{\mu\nu}(R+\frac{6}{l^2})\right)-T_{\mu\nu}=0,\tag{8c} 	
\end{align}
\setcounter{equation}{8}
where the stress-energy tensor is 
\begin{align}\label{SEtensor}
	T_{\mu\nu}=T^M_{\mu\nu}+T^1_{\mu\nu}+T^2_{\mu\nu},\tag{9a}
\end{align}
with the Maxwell and Dirac-field stress-energy tensors defined as follows:\begin{align}
	\label{SEtM}
	&T^M_{\mu\nu}=\mathcal{F}_{\mu\lambda}\mathcal{F}_{\nu\rho}g^{\lambda\rho}-\frac{1}{4}g_{\mu\nu}\mathcal{F}^2,\tag{9b}\\
	\label{SEtD}
	&T^{\epsilon}_{\mu\nu}=\text{Im}\left(\bar{\Psi}_{\epsilon}\gamma_{\mu}\hat{D}_{\nu}\Psi_{\epsilon}+\bar{\Psi}_{\epsilon}\gamma_{\nu}\hat{D}_{\mu}\Psi_{\epsilon}\right)\tag{9c}.
\end{align}
\setcounter{equation}{9}
We will solve EOM (\ref{EOM1})-(\ref{EOM3}) to get the traversable wormhole in AdS in the next section.

\section{Wormhole solutions}

By solving the Einstein-Dirac-Maxwell model (\ref{action}), we obtain the traversable wormhole solutions with spherical and planar topologies in an asymptotically AdS spacetime, respectively. We discuss them one by one below. 

\subsection{Spherical topology}

\begin{figure}[H]
	\centering
	\begin{minipage}[t]{0.32\textwidth}
		\centering
		\includegraphics[width=5.2cm]{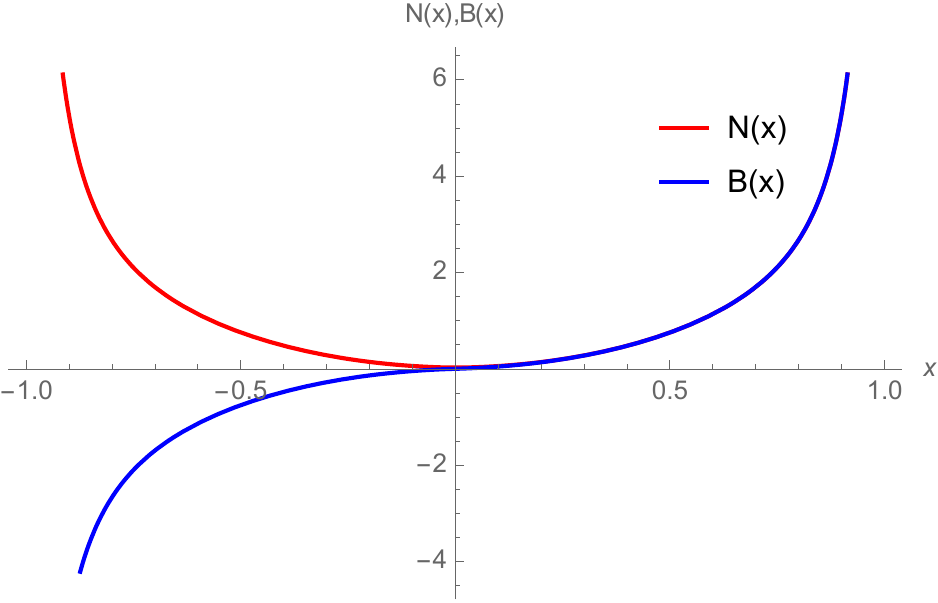}
	\end{minipage}
	\begin{minipage}[t]{0.32\textwidth}
		\centering
		\includegraphics[width=5.2cm]{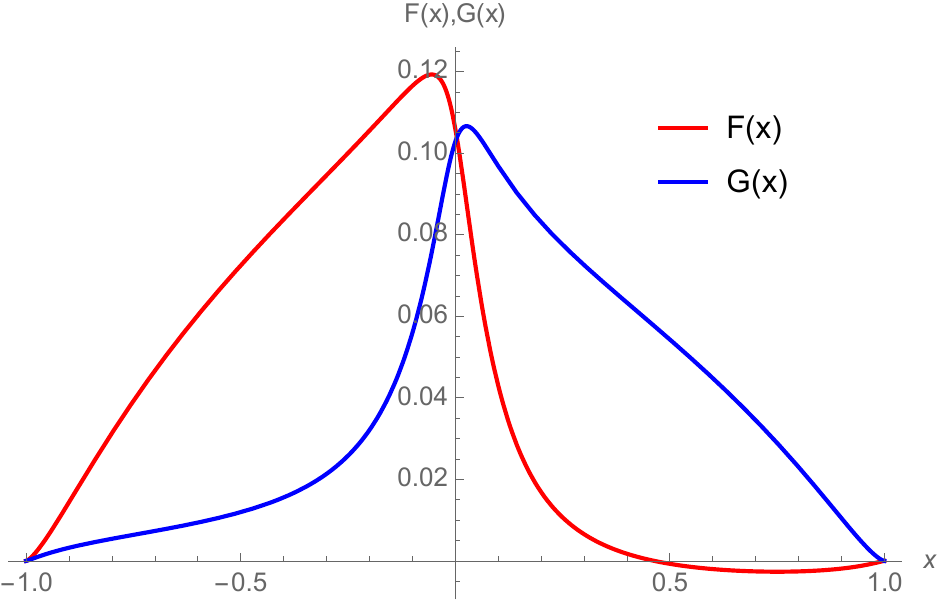}
	\end{minipage}
	\begin{minipage}[t]{0.32\textwidth}
		\centering
		\includegraphics[width=5.2cm]{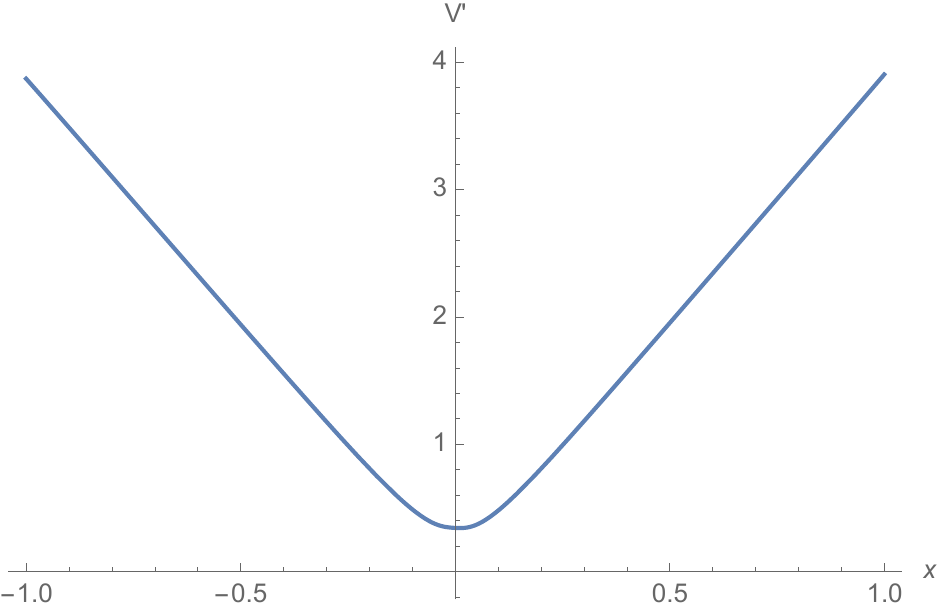}
	\end{minipage}
	\caption{Wormhole with spherical topology. The theory parameters and initial values are given by (\ref{paraS0}) and  (\ref{paraS}). }
	\label{fig:gwhsFIG}
\end{figure}

For the wormhole with a spherical topology, we take the following ansatz of metric 
\begin{align}\label{stWHga}
	ds^2=-N(r)^2dt^2+\frac{1}{B(r)^2}dr^2+r^2 d\Omega^2,
\end{align}
where $d\Omega^2$ denotes the line element of unit sphere and $r\ge r_0$ with $r_0$ the radius of wormhole throat. For the numerical convenience, we parameterize $r$ as \cite{Konoplya:2021hsm}
 \begin{align}
	\label{defr}r(x)=\frac{r_0}{1-x^2} ,\quad x=\pm\sqrt{1-\frac{r_0}{r}} .
\end{align}
Then, the metric becomes 
\begin{align}\label{manzatz}
	ds^2=-N(x)^2dt^2+\frac{r'(x)^2}{B(x)^2}dx^2+r(x)^2d\Omega^2. 
\end{align}
As $x$ goes from $-1\to 0 \to 1$, the radial coordinate $r(x)$ goes from $\infty \to r_0\to \infty$, which moves from one AdS boundary through the wormhole throat to the other AdS boundary. To have an asymptotically AdS spacetime, we impose the boundary conditions
\begin{align}\label{gadsC1}
	\lim_{x\rightarrow\pm1}&N(x)^2\to1+\frac{r^2}{l^2}\to 1+\frac{r_0^2}{l^2(1-x^2)^2} ,\tag{13a}\\
	\label{gadsC2}\lim_{x\rightarrow\pm 1}&B(x)^2\to 1+\frac{r^2}{l^2}\to 1+\frac{r_0^2}{l^2(1-x^2)^2},\tag{13b}
\end{align}
where a numerical cutoff of $x=\pm (1-\epsilon)$ should be taken. We take $\epsilon=10^{-4}$ in this section. 
Note that we have $\lim_{x\rightarrow\pm1}N(x)^2=\lim_{x\rightarrow\pm1}B(x)^2\to \infty$ for the AdS wormholes (see Fig.\ref{fig:gwhsFIG} ), while $\lim_{x\rightarrow\pm1}N(x)^2=\lim_{x\rightarrow\pm1}B(x)\to 1$ for the flat wormholes \cite{Blazquez-Salcedo:2020czn,Konoplya:2021hsm, Wang:2022aze}. This is the main difference between our wormhole solutions and those presented in \cite{Blazquez-Salcedo:2020czn, Konoplya:2021hsm, Wang:2022aze} for flat space.
\setcounter{equation}{13}

From metric (\ref{manzatz}), we read off the tetrad fields
\begin{align}\label{tetradS}
	e_{\mu a}=\begin{pmatrix}
		-N(x)&0&0&0\\
		0&\frac{r'(x)}{B(x)}&0&0\\
		0&0&r(x)&0\\
		0&0&0&r(x)\sin(\theta)
	\end{pmatrix} .
\end{align}
We employ the following ansatz for the vector 
\begin{align}
	\label{AA}
	A_{\mu}dx^{\mu}=V(x)dt\tag{15a},
\end{align}
and for the spinors
\begin{align}\label{gspinor1A}
	\Psi_1&=e^{-i\omega t+\frac{i\varphi}{2}}\begin{pmatrix}
		\phi(x)\cos\frac{\theta}{2}\\
		i\phi^*(x)\sin\frac{\theta}{2}\\
		-i\phi^*(x)\cos\frac{\theta}{2}\\
		-\phi(x)\sin\frac{\theta}{2}
	\end{pmatrix}\tag{15b}, \\
	\label{gspinor2A}
	\Psi_2&=e^{-i\omega t-\frac{i\varphi}{2}}\begin{pmatrix}
		i\phi(x)\sin\frac{\theta}{2}\\
		\phi^*(x)\cos\frac{\theta}{2}\\
		\phi^*(x)\sin\frac{\theta}{2}\\
		i\phi(x)\cos\frac{\theta}{2}
	\end{pmatrix}\tag{15c},
\end{align}
where \begin{align}\label{phiA}
	\phi(x)=F(x)e^{i\pi/4}-G(x)e^{-i\pi/4}\tag{15d} ,
\end{align}
with $F(x),G(x)$ real functions.
Substituting the above ansatzs into EOM \eqref{EOM1}-\eqref{EOM3}, we obtain a set of ordinary differential equations for $N(x),B(x),V(x),F(x)$ and $G(x)$. See Appendix A. These equations are first-order for $F(x)$, $G(x)$, $N(x)$  and $B(x)$, while second-order for $V(x)$. 
 \setcounter{equation}{15} 

\begin{figure}[H]
	\centering
	\begin{minipage}[t]{0.49\textwidth}
		\centering
		\includegraphics[width=6.2cm]{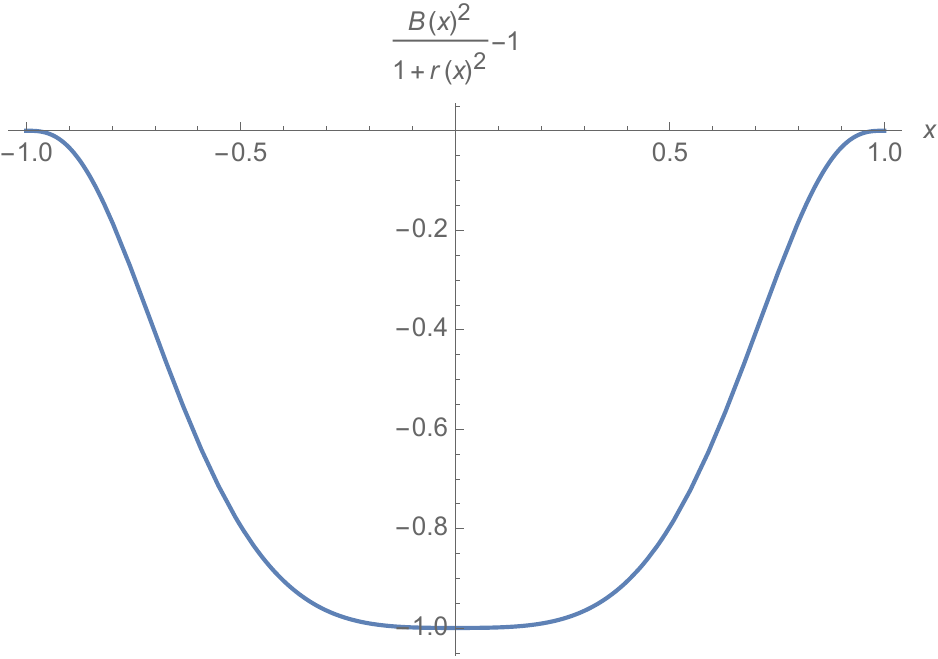}
	\end{minipage}
	\begin{minipage}[t]{0.49\textwidth}
		\centering
		\includegraphics[width=6.2cm]{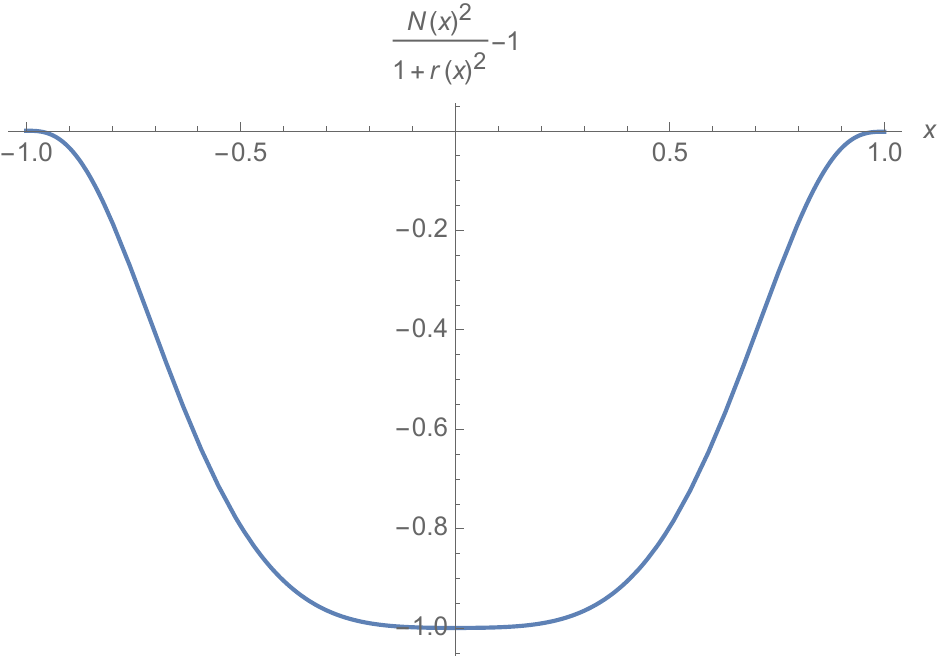}
	\end{minipage}
	\caption{The asymptotically-AdS conditions \eqref{gadsC1} and \eqref{gadsC2} are satisfied. }
	\label{fig:gwhs2FIG}
\end{figure}

We apply the shooting method to solve the wormhole solutions. To do so, we expand $A=(F,G,N,V,B)$  around the throat $x=0$,
 \begin{align}
	\label{seriesF}
	A(x)=a_0+a_1 x+O(x^2),
\end{align}  
which means $B(x)=b_0+b_1 x+O(x^2)$ and similar for other functions $(F,G,N,V)$. Solving EOMs perturbatively, we find the following constraints 
\begin{align}\label{near0f}
	b_0&=0,\\
	\omega &= -\frac{n_0  \left(b_1^2 + 16 m r_0^2 (f_0^2 - g_0^2)\right) - 32 n_0 r_0 f_0 g_0}{32 r_0^2  \left(f_0^2 + g_0^2\right)},\\
	v_1&=\sqrt{2}n_0 \sqrt{- \left(\frac{l^2  \left(b_1^2 - 2 + 16 m (g_0^2 - f_0^2) r_0^2\right) + 32 l^2 f_0 g_0 r_0 - 6  r_0^2}{l^2  b_1^2}\right)}.
\end{align}
For any given set of initial values (\ref{seriesF}), we can numerically solve the EOMs of Appendix A to obtain a solution. In general, the obtained solution does not obey the asymptotical boundary conditions (\ref{gadsC1},\ref{gadsC2}) at $|x|\to 1$. We adjust the initial values (\ref{seriesF}) to satisfy the boundary conditions (\ref{gadsC1},\ref{gadsC2}). This is the so-called shooting method. Without loss of generality, we choose the  theory parameters 
\begin{align}\label{paraS0}
	r_0 = 1, \quad l = 1, \quad m = 0.2, \quad q = 0.03,  
	\end{align}
and find the following initial values can do the work
\begin{align}\label{paraS}
\quad v_0 = 0, \quad n_0 = 0.025, \quad b_1 = 0.29, \quad f_0=0.106, \quad g_0=0.103.
\end{align}
The corresponding functions for the wormhole are drawn in Fig. \ref{fig:gwhsFIG}.  As shown in Fig. \ref{fig:gwhs2FIG}, the wormhole solution indeed obeys the boundary conditions (\ref{gadsC1},\ref{gadsC2}) on the asymptotically AdS boundary $|x|\to 1$. For instance, we have numerically
\begin{align}\label{WHBCs1}
	&\left. \frac{B(x)^2}{1+r(x)^2}-1 \right|_-=2.31\times 10^{-9}  ,
	\left. \frac{B(x)^2}{1+r(x)^2}-1 \right|_+=2.04\times 10^{-9} ,\\
	&\left. \frac{N(x)^2}{1+r(x)^2}-1 \right|_-=8.50\times10^{-5},\quad \left. \frac{N(x)^2}{1+r(x)^2}-1 \right|_+=-0.0019.\label{WHBCs2}
\end{align}
on the AdS boundary $|x|=1-10^{-4}$. Note that the last term of (\ref{WHBCs2}) does not vanishes perfectly. According to \cite{Blazquez-Salcedo:2020czn}, it implies that there is a slight redshift on the right AdS boundary. 


\begin{figure}[H]
	\centering
	\includegraphics[width=0.6\linewidth]{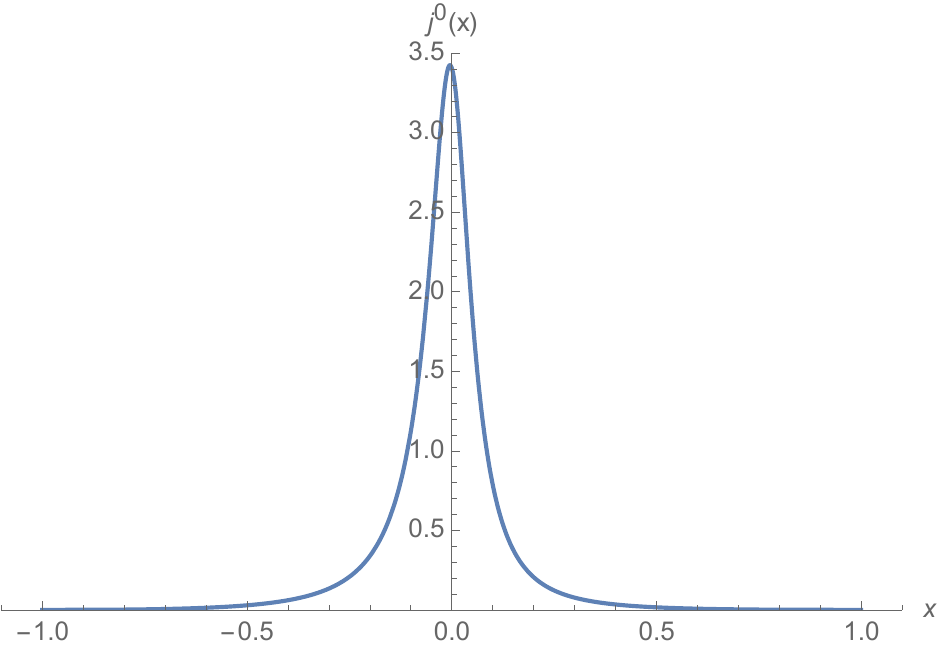}
	\caption{Charge density of spherical wormhole. The charge is distributed throughout the entire space (consistent with Pauli exclusion principle), reaching a maximum near the throat ($x=0$) and gradually approaching zero at the AdS boundary ($|x|=1$).}
	\label{fig:J0}
\end{figure}

To examine the fermion distributions, we calculate the charge density given by 
\begin{align}
	j^0=\sum_{\epsilon=1,2}\bar{\Psi}_{\epsilon}\gamma^0 \Psi_{\epsilon}=\frac{4\left(F(x)^2+G(x)^2\right)}{ N(x)},
\end{align}
where the result is illustrated in Fig. \ref{fig:J0}. The plot indicates that the charge is distributed throughout the entire space, reaching a maximum near the throat ($x=0$) and gradually approaching zero at the AdS boundary ($|x|=1$).

\begin{figure}[H]
	\centering
	\includegraphics[width=0.6\linewidth]{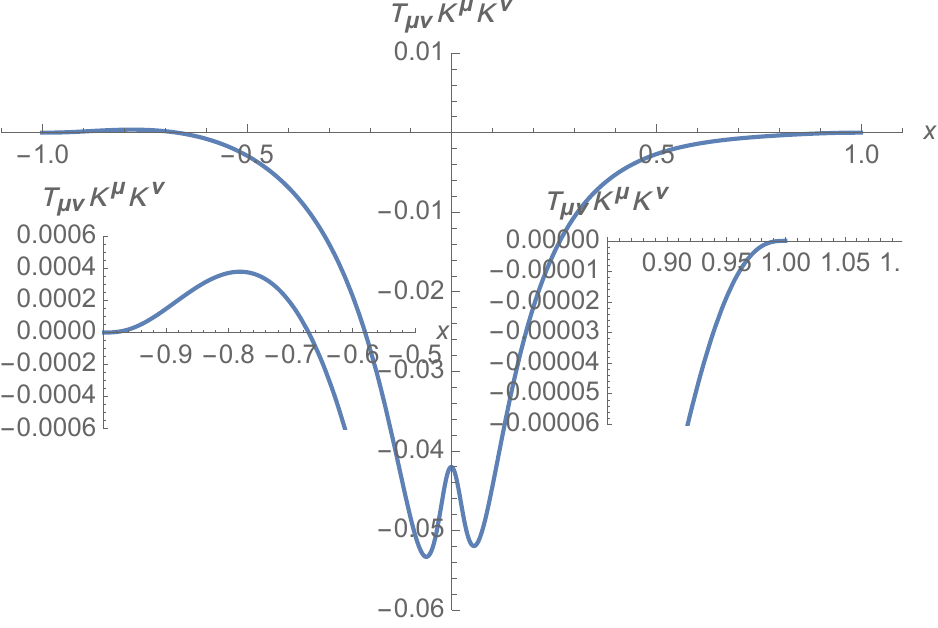}
	\caption{Violation of the NEC for the wormhole with spherical topology, i.e., $T_{\mu\nu}K^{\mu}K^{\nu}<0$. }
	\label{fig:gwhTnull}
\end{figure}

It is well-known that a traversable wormhole violates the null energy condition (NEC). For the null vector
\begin{align}\label{nullVS}
	K^{\mu}\partial_{\mu} = N(x)^{-1} \partial_t + \frac{B(x)}{r'(x)} \partial_x,
\end{align}
we verify that the NEC is violated for our wormhole solution, i.e., $T_{\mu\nu}K^{\mu}K^{\nu}<0$. See Fig.  \ref{fig:gwhTnull}. Thus, we indeed obtain a traversable wormhole in the asymptotically AdS spacetime. 
Fig. \ref{fig:gwhTnull} shows NEC is most severely violated near the throat of the wormhole $x\to 0$, while it is essentially not violated near the AdS boundary $x\to \pm 1$. More precisely, we numerically observe the NEC is violated in the bulk region $ -0.67<x<1$.

\subsection{Planar topology}

\begin{figure}[H]
	\centering
	\begin{minipage}[t]{0.42\textwidth}
		\centering
		\includegraphics[width=6.2cm]{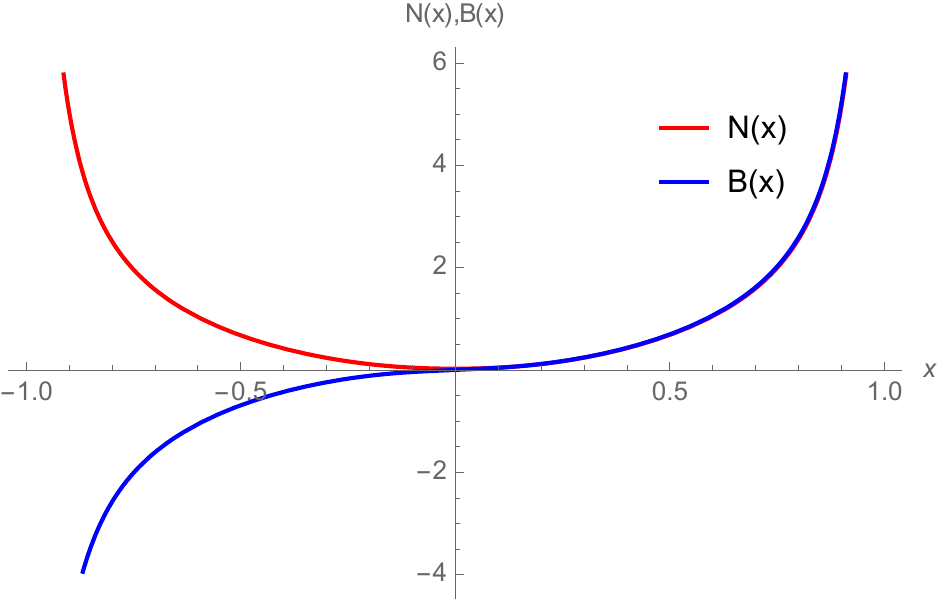}
	\end{minipage}
	\begin{minipage}[t]{0.42\textwidth}
		\centering
		\includegraphics[width=6.2cm]{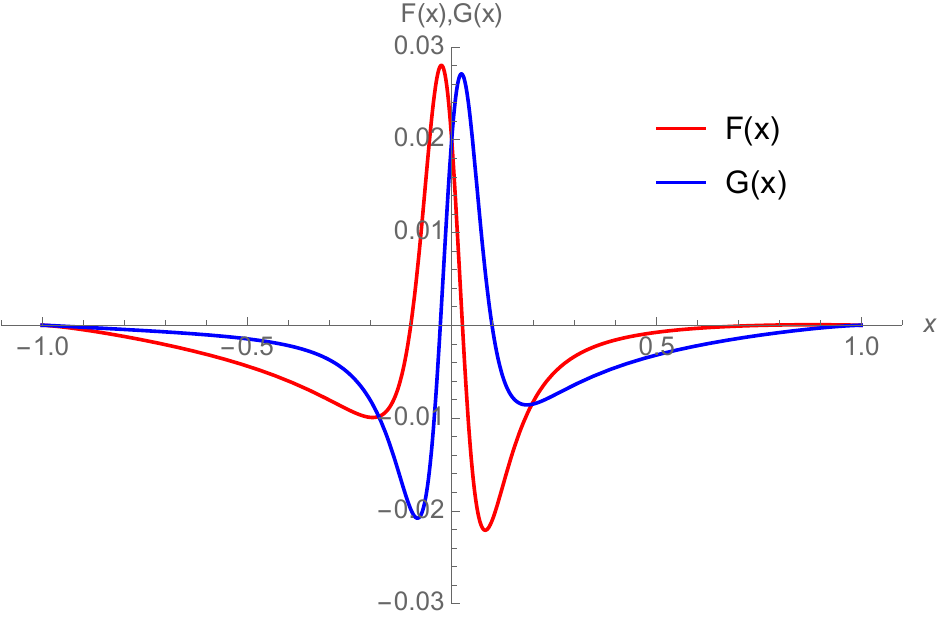}
	\end{minipage}\\
	\begin{minipage}[t]{0.42\textwidth}
		\centering
		\includegraphics[width=6.2cm]{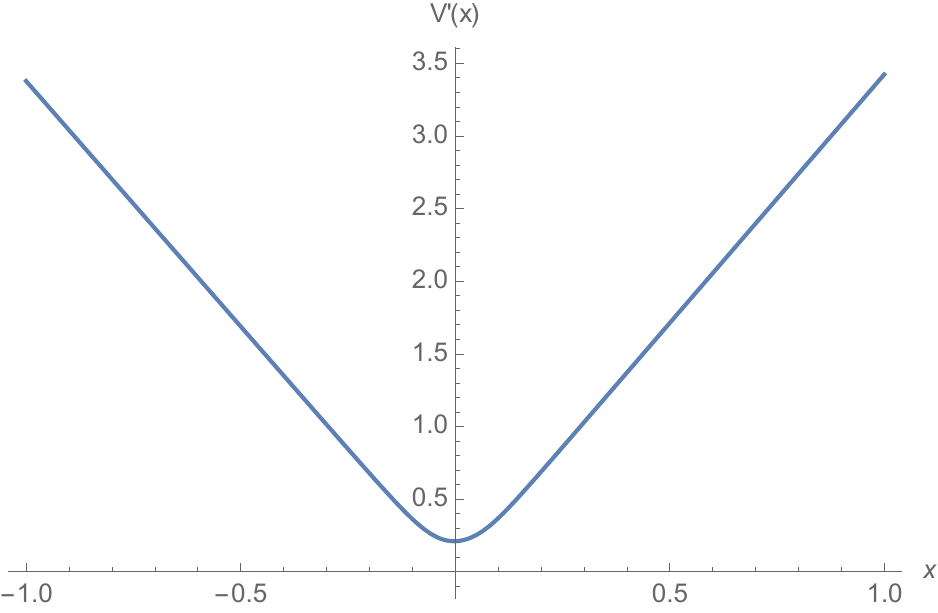}
	\end{minipage}
		\begin{minipage}[t]{0.42\textwidth}
		\centering
		\includegraphics[width=6.2cm]{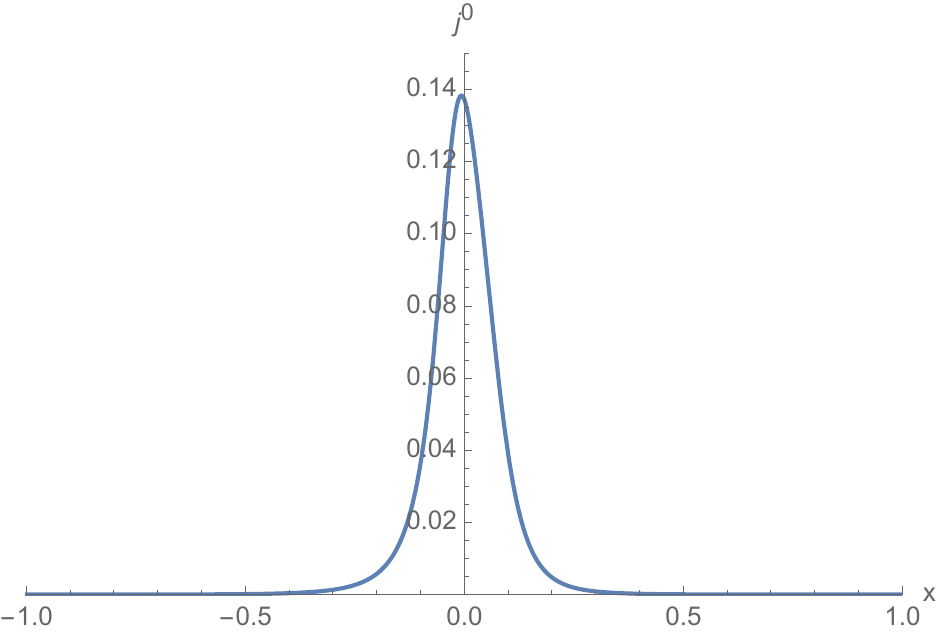}
	\end{minipage}
	\caption{The planar wormhole and charge density with parameters (\ref{parap}).}
	\label{fig:pwhsFIG}
\end{figure}

\begin{figure}[H]
	\centering
	\includegraphics[width=0.7\linewidth]{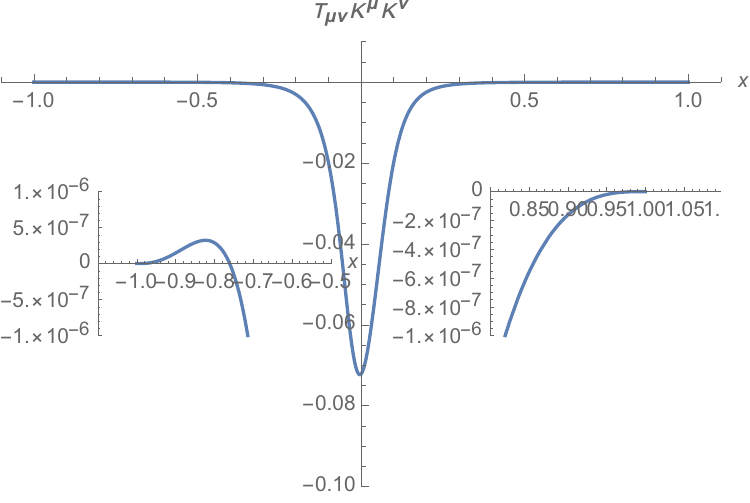}
	\caption{Violation of the NEC for the wormhole with planar topology, i.e., $T_{\mu\nu}K^{\mu}K^{\nu}<0$.}
	\label{fig:pwhTnull}
\end{figure}

Let us go on to study the wormhole with planar topology. Since the calculations are similar to those of spherical topology, we show only key points below. We take the ansatz of metric
\begin{align}
	\label{pwhA}
	ds^2=-N(x)^2dt^2+\frac{r'(x)^2}{B(x)^2}dx^2+r^2(dy_1^2+dy_2^2),
\end{align}
where we have set the AdS radius $l=1$ for simplicity. We require the (\ref{pwhA}) is an asymptotically-AdS metric
\begin{align}\label{pwhadsC1}
\lim_{x\rightarrow\pm 1}N(x)^2\to r^2 ,\tag{27a}\\
\lim_{x\rightarrow\pm 1}B(x)^2\to r^2 .\tag{27b} \label{pwhadsC2}
\end{align} \setcounter{equation}{27}
From (\ref{pwhA}), we get the tetrads 
\begin{align}\label{tetradP}
	e_{\mu a}=\begin{pmatrix}
		-N(x)&0&0&0\\
		0&\frac{r'(x)}{B(x)}&0&0\\
		0&0&r(x)&0\\
		0&0&0&r(x)
	\end{pmatrix} .
\end{align}
The ansatz for vector and spinors are given by (\ref{AA}) and
\begin{align}
	\label{pspinor1}&\Psi_1=e^{-i\omega t} 
	\begin{pmatrix}
		\phi(r)\\
		0\\
		-i \phi^*(r)\\
		0
	\end{pmatrix}\tag{29a} ,\\
	\label{pspinor2}&\Psi_2=e^{-i\omega t} 
	\begin{pmatrix}
		0\\
		\phi^*(r)\\
		0\\
		i\phi(r)
	\end{pmatrix}\tag{29b}.
\end{align}
\setcounter{equation}{29}
Substituting \eqref{AA}, \eqref{phiA}, \eqref{pspinor1} and \eqref{pspinor2}
into EOMs \eqref{EOM1}-\eqref{EOM3}, we obtain a set of ordinary differential equations for $N(x),B(x),V(x),F(x)$ and $G(x)$. See Appendix B. Solving these EOMs perturbatively around the throat $x=0$, we derive
\begin{align}\label{pnear0f}
	b_0 &= 0 , \\
	\omega &= -\frac{n_0(b_1^2 + 16 (f_0^2 - g_0^2) m r_0^2)}{32(f_0^2 + g_0^2) r_0^2} , \\
	v_1 &= \sqrt{\frac{2 n_0^2 \left(-b_1^2 + 2r_0^2 \left(3 + 8 (f_0^2 - g_0^2) m \right)\right)}{b_1^2}} .
\end{align}
We choose the parameters and fix the initial values by using the shooting method 
 \begin{align}\label{parap}
	r_0=1,\ m=0.2,\ q=0.03,\  v_0=0,\ n_0=0.38,\ b_1=0.023,\ f_0=g_0=0.02. 
\end{align}
We draw one typical solution of the planar wormhole and the charge density in Fig. \ref{fig:pwhsFIG}. We also verify the violation of the NEC in Fig. \ref{fig:pwhTnull}, i.e., $T_{\mu\nu}K^{\mu}K^{\nu}<0$. 
We numerically find the NEC is violated in the bulk region  $ -0.76<x<1$ . The closer to the wormhole throat, the more severely the NEC is violated. 

\begin{figure}[H]
	\centering
	\begin{minipage}[t]{0.49\textwidth}
		\centering
		\includegraphics[width=6.2cm]{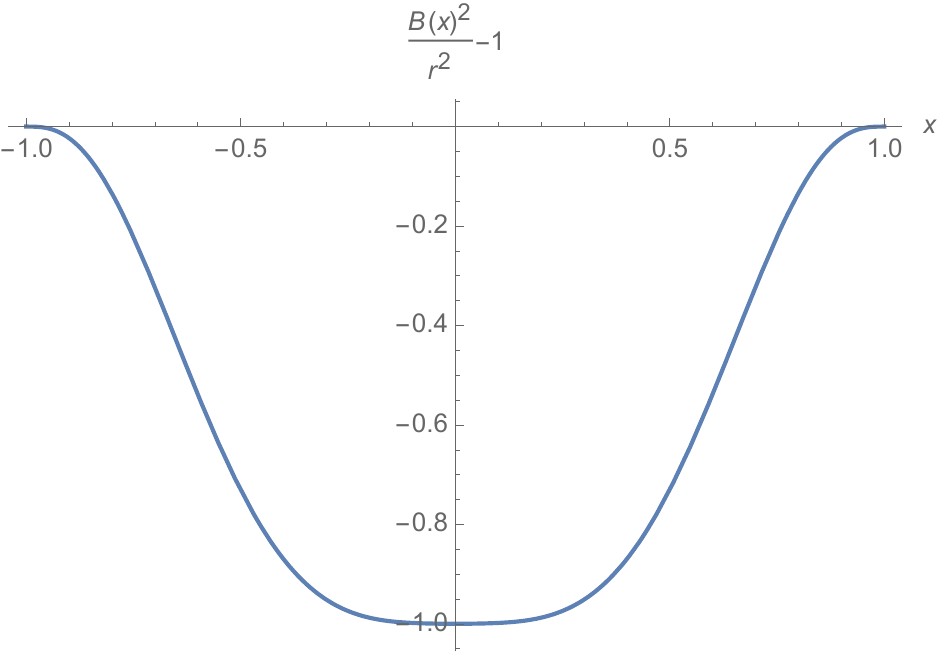}
	\end{minipage}
	\begin{minipage}[t]{0.49\textwidth}
		\centering
		\includegraphics[width=6.2cm]{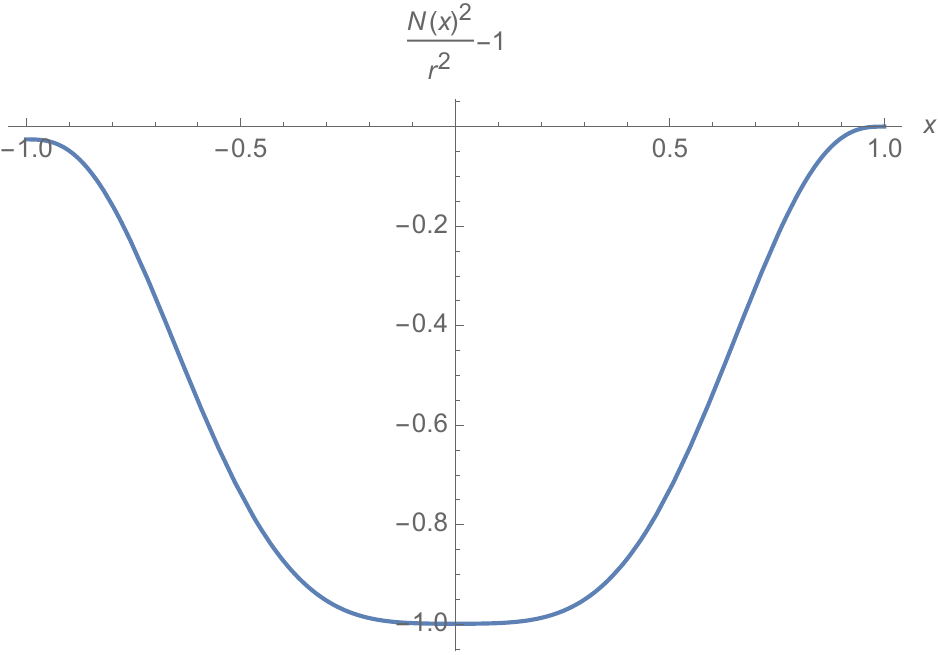}
	\end{minipage}
	\caption{The asymptotically-AdS conditions (\ref{pwhadsC1},\ref{pwhadsC2}) are satisfied up to a redshift.}
	\label{fig:pwhbadsp}
\end{figure}

As shown in Figure \ref{fig:pwhbadsp}, the wormhole solution obeys well the boundary conditions on the asymptotically AdS boundary $|x|\to 1$. For instance, we have numerically
\begin{align}\label{PWHBCs}
	&\left. \frac{B(x)^2}{r(x)^2}-1 \right|_-=-3.14\times 10^{-11}  ,
	\left. \frac{B(x)^2}{r(x)^2}-1 \right|_+=-3.14\times 10^{-11} ,\\
	&\left. \frac{N(x)^2}{r(x)^2}-1 \right|_-=-0.026,\quad \left. \frac{N(x)^2}{r(x)^2}-1 \right|_+=-1.10\times 10^{-6},
\end{align}
where have redefined time as $t\rightarrow 0.99 t $ to make $(N^2/r^2)|_+\to 1$ on the right AdS boundary. Then, we have $(N^2/r^2)|_-\approx 0.97$ on the left AdS boundary, which can be explained as a redshift \cite{Blazquez-Salcedo:2020czn}. 

\label{subsec:pwhsolution}

\section{HEE of strip}

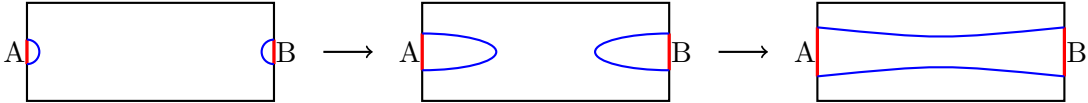
\begin{figure}[H]
	\begin{tikzpicture}[scale=1]
		\draw[thick] (0,0) rectangle (3.25,1.3);
		
		\draw[very thick, red] (0,0.4875) -- (0,0.8125);
		\draw[very thick, red] (3.25,0.4875) -- (3.25,0.8125);
		
		\draw[blue, thick] (0,0.8125) arc[start angle=90, end angle=-90, radius=0.1625];
		\draw[blue, thick] (3.25,0.8125) arc[start angle=90, end angle=270, radius=0.1625];
		
		\node at (-0.1625,0.65) {A};
		\node at (3.4125,0.65) {B};
		
		\draw[thick, ->] (3.9,0.65) -- (4.55,0.65);
		
		\begin{scope}[shift={(5.2,0)}]
			\draw[thick] (0,0) rectangle (3.25,1.3);
			
			\draw[very thick, red] (0,0.40625) -- (0,0.89375);
			\draw[very thick, red] (3.25,0.40625) -- (3.25,0.89375);
			
			\draw[blue, thick] (0,0.89375) arc[start angle=90, end angle=-90, x radius=0.975, y radius=0.24375];
			\draw[blue, thick] (3.25,0.89375) arc[start angle=90, end angle=270, x radius=0.975, y radius=0.24375];
			
			\node at (-0.1625,0.65) {A};
			\node at (3.4125,0.65) {B};
			
			\draw[thick, ->] (3.9,0.65) -- (4.55,0.65);
		\end{scope}
		
		\begin{scope}[shift={(10.4,0)}]
			\draw[thick] (0,0) rectangle (3.25,1.3);
			
			\draw[very thick, red] (0,0.325) -- (0,0.975);
			\draw[very thick, red] (3.25,0.325) -- (3.25,0.975);
			
			\draw[blue, thick] (0,0.975) .. controls (1.625,0.8125) .. (3.25,0.975);
			\draw[blue, thick] (0,0.325) .. controls (1.625,0.4875) .. (3.25,0.325);
			
			\node at (-0.1625,0.65) {A};
			\node at (3.4125,0.65) {B};
		\end{scope}
	\end{tikzpicture}
	\caption{Phase transition of RT surface.  As the strip width (red line segment) increases, the RT surface (blue curve) transforms from a disconnected phase (left) to the connected
	 phase (right).}
	\label{fig:changeA}
\end{figure}

This section investigates the holographic entanglement entropy (HEE) \cite{Ryu:2006bv} of two strips on the two AdS boundaries of the  traversable wormhole. For simplicity, we focus on the planar topology. As shown in Fig. \ref{fig:changeA}, there are two phases for the RT surfaces: one is connected and the other is disconnected.  

Recall the HEE can be calculated by the area of the minimal surface (RT surface) in bulk \cite{Ryu:2006bv}
\begin{align}\label{RTformula}
	S_A=\frac{\text{Area of } \gamma_A}{4} ,
\end{align} 
where $A$ denotes the subsystem on the AdS boundary, $\gamma_A$ is the minimal surface in bulk whose boundary coincides with that of $A$, i.e., $\partial \gamma_A=\partial A$. For the strip $-L/2\le y_1\le L/2$, the embedding function of RT surface in bulk is assumed to be
\begin{align}\label{esA}
	t=\text{constant}, \ y_1=y_1(x).
\end{align} 
Then, we derive the induced metric of the RT surface
\begin{align}\label{dM}
	\hat{ds}^2=\Big( \frac{r'(x)^2}{B(x)^2}+r(x)^2y_1'(x)^2 \Big) dx^2+r(x)^2dy_2^2 ,
\end{align} 
with the area
\begin{align}\label{Aformula}
	\mathcal{A}=\int dx\mathcal{L}=\int dx \ r(x)\sqrt{r(x)^2y_1'(x)^2+\frac{r'(x)^2}{B(x)^2}},
\end{align}
where we have set $\int dy_2=1$. Since $\mathcal{L}$ includes no $y(x)$, we can define a conserved quantity
\begin{align}\label{EOMAs}
	E=\frac{\partial\mathcal{L}}{\partial y_1'}=\frac{r_0^2 y_1'}{(1-x^2)\sqrt{y_1'^2(1-x^2)^2+\frac{4x^2}{B(x)^2}}}=\text{constant}.
\end{align}

\begin{figure}[H]
	\centering
	\includegraphics[width=0.5\linewidth]{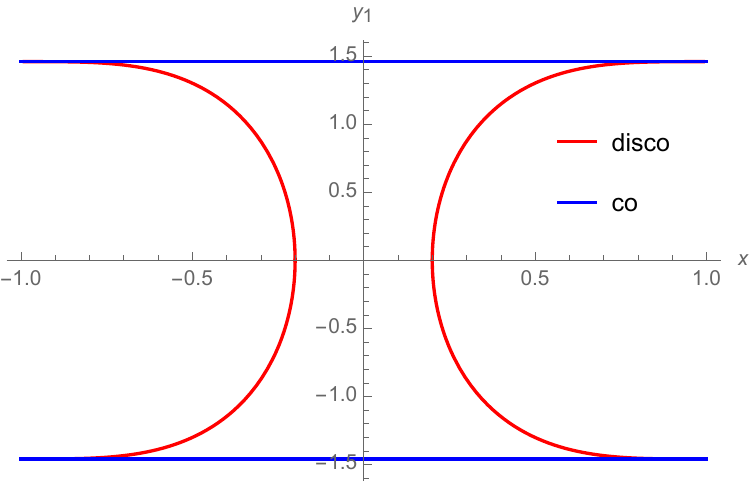}
	\caption{Two types of extremal surfaces with strip width $L=1.48$. The red and blue curves correspond to the disconnected and connected phases, respectively.}
	\label{fig:stripES}
\end{figure}

Let us first discuss the disconnected phase (red curve of Fig. \ref{fig:stripES}), which is dominated for the small strip width.
As shown in Fig. \ref{fig:changeA} and Fig. \ref{fig:stripES}, there is a turning point $x_{\text{min}}$ for the disconnected phase. At the turning point, we have 
\begin{align}
	\label{bdyC1}
	\quad\text{disconnect phase}:\quad y_1(x_{\text{min}})=0,\quad y'_1(x_{\text{min}})=\infty. 
\end{align}
Substituting it into (\ref{EOMAs}), we get
\begin{align}\label{EOMAs1}
	E=\frac{r_0^2}{\left(1-x_{\text{min}}^2\right)^2} 
\end{align}
On the right hand side of wormhole ($x>0$), from (\ref{EOMAs}) and (\ref{EOMAs1}), we solve $y'_1(x)$ and obtain the strip width
\begin{align}\label{strip width discon}
L&=2 \int_{x_{\min}}^{1-\epsilon} y_1'(x) dx\nonumber\\
&=2 \int_{x_{\min}}^{1-\epsilon}\frac{2 x \left(1-x^2\right) \   dx }{B(x) \sqrt{x_{\min }^8-4 x_{\min }^6+6 x_{\min }^4-4 x_{\min }^2-x^8+4 x^6-6 x^4+4 x^2}}.
\end{align}
We choose the same strip width $L$ on the left hand side of wormhole ($x<0$). Note that $|x_{\min}|$ are slightly different on the two sides of wormhole since  $|B(x)|$ are not exactly symmetric. 

Substituting $y'(x)$ into (\ref{Aformula}), we get the area of extremal surface $\mathcal{A}_{\text{disco}}=\mathcal{A}_{\text{disco},x>0}+\mathcal{A}_{\text{disco},x<0}$ with
\begin{align}\label{Aformula1}
	\mathcal{A}_{\text{disco},x>0}&=2\int_{x_{\min}}^{1-\epsilon} \frac{2 r_0^2 x \left(1-x_{\min }^2\right){}^2  \ \ dx}{\left(1-x^2\right)^3 B(x) \sqrt{x_{\min }^8-4 x_{\min }^6+6 x_{\min }^4-4 x_{\min }^2-x^8+4 x^6-6 x^4+4 x^2}},\\
	\label{Aformula1a}
	\mathcal{A}_{\text{disco},x<0}&=2\int^{x_{\min}}_{-1+\epsilon} \frac{2 r_0^2 x \left(1-x_{\min }^2\right){}^2  \ \ dx}{\left(1-x^2\right)^3 B(x) \sqrt{x_{\min }^8-4 x_{\min }^6+6 x_{\min }^4-4 x_{\min }^2-x^8+4 x^6-6 x^4+4 x^2}},
\end{align}
where recall that $|x_{\min}|$ are not exactly the same on the two sides of wormhole. 

Let us go on to discuss the connected phase (blue curve of Fig. \ref{fig:stripES}), which is dominated for the large strip width.  From (\ref{EOMAs}), we observe the solution for the connected phase is $y_1'(x)=0$. Then, the area of extremal surface becomes
\begin{align}\label{Aformula2}
	\mathcal{A}_{\text{co}}=2\int_{-1+\epsilon}^{1-\epsilon} \frac{2 r_0^2 x\ \ dx}{\left(1-x^2\right)^3 B(x)}.
\end{align}
which is a constant for fixed wormhole geometry. 

\begin{figure}[H]
	\centering
	\includegraphics[width=0.7\linewidth]{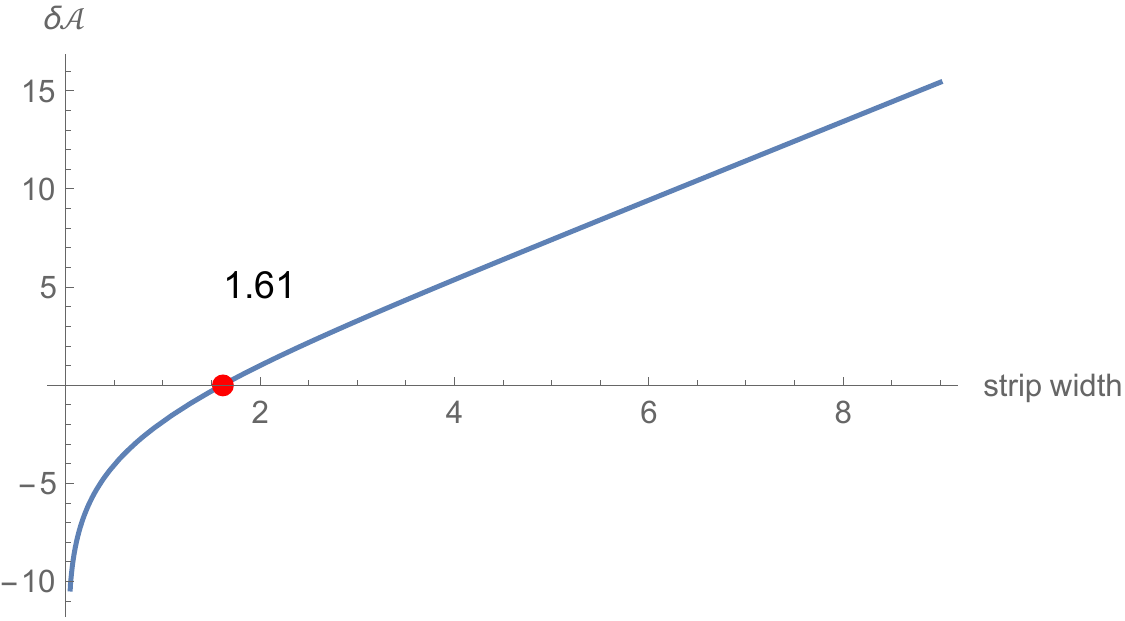}
	\caption{ The area difference $\delta \mathcal{A}=\mathcal{A}_{\text{disco}}-\mathcal{A}_{\text{co}}$ increases with the strip width $L$. The disconnected and connected phases dominate for $L<1.61$ and $L>1.61$, respectively.  }
	\label{fig:stripDA}
\end{figure}

By definition, the RT surface is given by the extremal surface with minimal area. We remark that, as $x_{\min}$ decreases, both the strip width $L$ (\ref{strip width discon}) and the area of extremal surface $\mathcal{A}_{\text{disco}}$ (\ref{Aformula1},\ref{Aformula1a}) increase. Thus,  for sufficiently large $L$,  $\mathcal{A}_{\text{disco}}$ could be larger than $\mathcal{A}_{\text{co}}$. Then, there is a phase transition and the connected phase becomes dominated.  We draw the $\delta \mathcal{A}-L$ relation in Fig. \ref{fig:stripDA}, where $\delta \mathcal{A}=\mathcal{A}_{\text{disco}}-\mathcal{A}_{\text{co}}$. It shows the disconnected phase dominates $\delta \mathcal{A}<0$ for small $L$, while the connected phase dominates $\delta \mathcal{A}>0$ for large  $L$. A phase transition of RT surface occurs near the strip width $L\approx 1.61$.


To end this section, let us comment on the possible relation between the NEC and the phase transition of holographic entanglement entropy. Recall that the NEC is most severely violated near the wormhole's throat. Similarly, the HEE undergoes a phase transition when the RT surface tends deeply into the bulk region near the wormhole throat. Thus, it is natural to ask if violating the null energy condition causes the phase transition of holographic entanglement entropy. It is qualitatively correct. For instance, the turning point of the disconnected extremal surface at the phase-transition point is $\left|x_{\text{min}}\right|\approx 0.37$, localized in the region violating the NEC. However, we do not find an exact relation, if any, between the NEC and the phase transition of HEE. It is beyond the primary purpose of this paper. We leave this interesting open question for future work.

\section{HEE of disk}

This section discusses the HEE of disks on the two AdS boundaries of the planar wormhole. The new feature is that the connected extremal surface disappears for sufficiently small disk radius. It is similar to the case of eternal black hole \cite{Hu:2022zgy}, where the Hartman-Maldacena (HM) surface of a disk disappears for sufficiently long time. On the other hand, for a strip, there are always connected extremal surface for a wormhole and HM surface for an eternal black hole. We first illustrate this unusual situation in an analytical toy model and then generalize the results to our numerical wormhole.

\subsection{A toy model}

\begin{figure}[H]
 	\centering
	 \begin{minipage}[t]{0.48\textwidth}
 		\centering
 		\includegraphics[width=6.2cm]{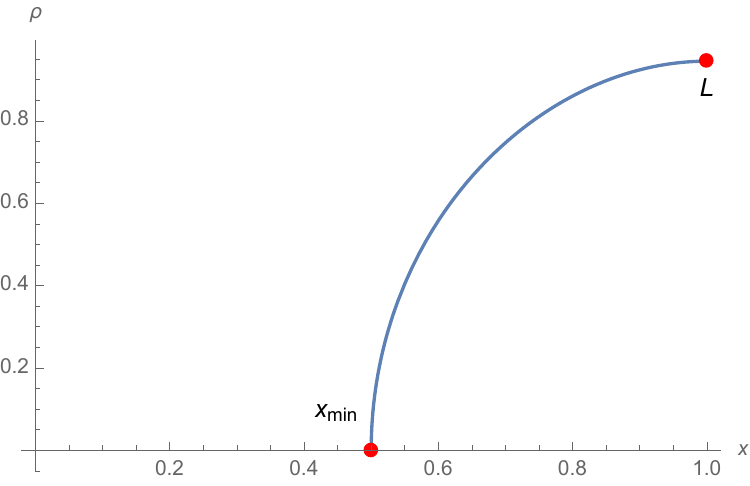}
 	\end{minipage}
 	\begin{minipage}[t]{0.48\textwidth}
 		\centering
 		\includegraphics[width=6.2cm]{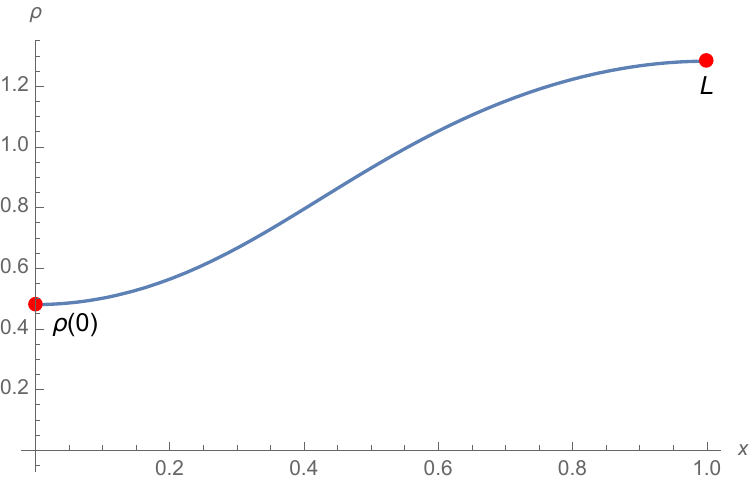}
 	\end{minipage}
 	\caption{The disconnected (left) and connected (right) extremal surfaces of disks for $x > 0$.  Here  $x_{\text{min}}$ is the turning point of the disconnected extremal surface and $L=\rho(1)$ is the disk radius.}
 	\label{fig:diskES1}
 \end{figure}

\begin{figure}[H]
	\centering
	\begin{minipage}[t]{0.7\textwidth}
		\centering
		\includegraphics[width=8cm]{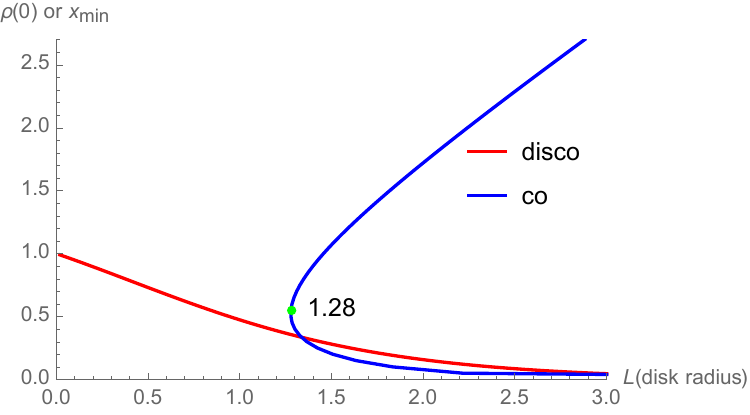}
	\end{minipage}
	\caption{ The figure shows the disk radius has a minimal value (green point) for the connected phase. Besides, one disk radius corresponds to one disconnected surface and two connected surfaces generally.}
	\label{fig:diskES2}
\end{figure}

We start with a toy model with the metric	
\begin{align}\label{tgA}
	ds^2=-(\tilde{r}^2+a^2)dt^2+\frac{1}{\tilde{r}^2+a^2}d\tilde{r}^2+(\tilde{r}^2+r_0^2)(d\rho^2+\rho^2d\theta^2),
\end{align}
which violates the NEC
\begin{align}\label{tvNEC}
	T_{\mu\nu}K^{\mu}K^{\nu}=-\frac{2r_0^2(\tilde{r}^2+a^2)^2}{(\tilde{r}^2+r_0^2)^2}<0,\quad K^{\mu}=(1,\tilde{r}^2+a^2,0,0),
\end{align}
and approaches $\text{Poincar}\acute{\text{e}}$-AdS for $\tilde{r}\rightarrow \infty$. Thus, it is a wormhole in an  asymptotically AdS space. Note that it is not a solution to the Einstein-Dirac-Maxwell model (\ref{action}). We just take it to illustrate the unusual situation of HEE for a disk with $\rho\le L$. Similar to sect. 2, we define 
\begin{align}\label{trrp}
	r^2=\tilde{r}^2+r_0^2,\quad r=r(x)=\frac{r_0}{1-x^2},
\end{align}
and assume the embedding function of RT surface in bulk 
\begin{align}\label{esA}
	t=\text{constant}, \ \rho=\rho(x).
\end{align} 
Then, the induced metric on the RT surface reads
\begin{align}\label{tgA2}
	\hat{ds}^2&=\left(g_{xx}(x)+r(x)^2\rho'(x)^2\right)dx^2+r(x)^2\rho(x)^2d\theta ^2,\\
	g_{xx}(x)&=\frac{r'(x)^2}{r(x)^2+a^2-2r_0^2+\frac{r_0^4-r_0^2a^2}{r(x)^2}}.
\end{align}
From (\ref{tgA2}), we derive the area functional of RT surface
\begin{align}
	\label{tformulaA2}
	\mathcal{A}&=4\pi\int_{ 0, x_{\min}}^{1-\epsilon} dx\ r(x)\rho(x)\sqrt{g_{xx}(x)+r(x)^2\rho '(x)^2}, 
\end{align}
where the integral are performed on one side of the wormhole and we have added a factor $2$ to account the whole space. The above area functional yields the Euler-Lagrange equation
\begin{align}
	\label{eomdisk}
	\rho''(x)-\frac{(1 - x^2)^2 g_{xx}(x)}{r_0^2 \rho(x)} +\frac{6 x \rho'(x)}{1 - x^2} - \frac{\rho'(x)^2}{\rho(x)} +\frac{\rho'(x) \left(8 r_0^2 x \rho'(x)^2-(1 - x^2)^3 g_{xx}'(x)\right)}{2 (1 - x^2)^3 g_{xx}(x)}=0.
\end{align}
Similar to the case of strip, there are disconnected and connected extremal surfaces for the disk, where the corresponding boundary conditions read
 \begin{align}
	\label{bdyDC1}
	\quad\text{disconnected phase}:&\quad \rho(x_{\text{min}})=0,\quad \rho'(x_{\text{min}})=\infty,\\
	\label{bdyDC2}
	\quad\text{connected phase}:&\quad \rho'(0)=0.
\end{align}
Here $x_{\text{min}}$ is the turning point of disconnected extremal surfaces. See Fig. \ref{fig:diskES1} for example. Without loss of generality, we set $a=r_0=1$ below.  We remark that the connected phase occurs only if the disk radius is larger than a critical value  $L\ge 1.28$, shown as the green point of Fig. \ref{fig:diskES2}. Besides, Fig. \ref{fig:diskES2} shows one disk radius corresponds to two or three different extremal surfaces generally.  We choose the one with minimal area as the RT surface.

\begin{figure}[H]
		\centering
		\includegraphics[width=10cm]{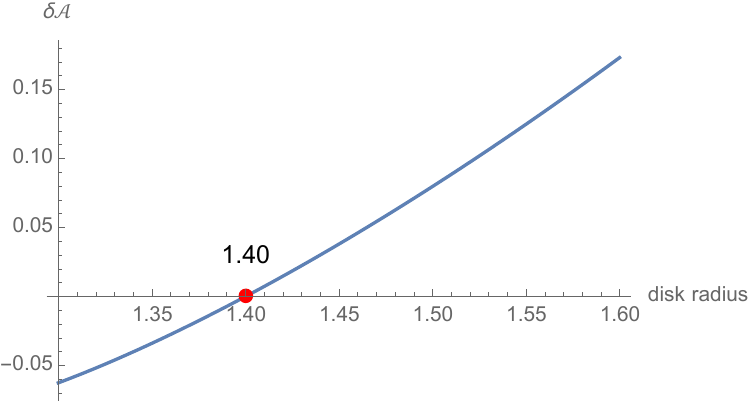}
	\caption{Existence of phase transition(red point); The area difference $\delta \mathcal{A}=\mathcal{A}_{\text{disco}}-\mathcal{A}_{\text{co}}$ increases with the disk radius $L$. The disconnected and connected phases dominate for $L<1.40$ and $L>1.40$, respectively.}
	\label{fig:diskDA}
\end{figure}

We draw the area difference  $\delta \mathcal{A}=\mathcal{A}_{\text{disco}}-\mathcal{A}_{\text{co}}$ in Fig. \ref{fig:diskDA}. It shows the area difference $\delta \mathcal{A}$ increases with the disk radius $L$. The disconnected and connected phases dominate for $L<1.40$ and $L>1.40$, respectively. Note that, the critical point $L_c\approx 1.28$ for the existence of connected phase is smaller than the phase-transition point $L_p\approx 1.40$. It aligns with the expectation that for a sufficiently large disk size, the connected phase invariably dominates. In such a scenario, the extremal surface would traverse through the wormhole throat.

\subsection{Planar wormhole}

\begin{figure}[H]
	\centering
	\includegraphics[width=0.6\linewidth]{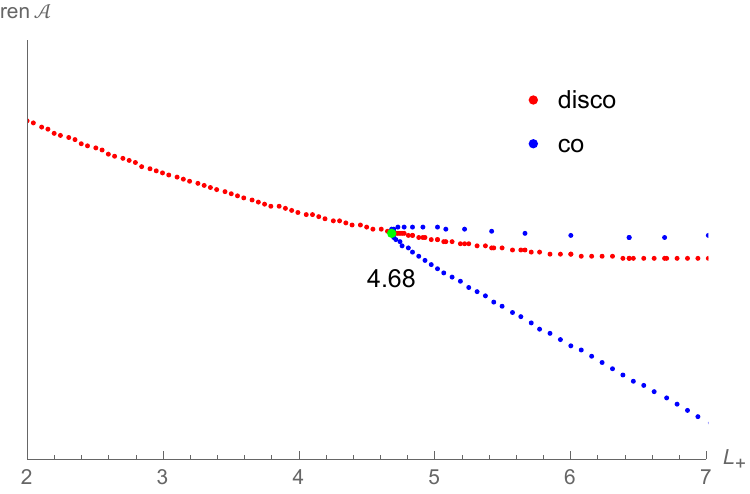}
	\caption{ Relation between the renormalized area $( \mathcal{A}-\frac{L}{\epsilon})$ and the right disk radius $L_+$. For $L_+< L_c\approx  4.68$, only the disconnected phase exists and we choose $L_-=L_+$. While for $L_+>L_c\approx  4.68$, $L_-$ can be determined by $L_+$ in the connected phase, which is slightly different from $L_+$ due to the slight asymmetry of the two sides of wormhole. See Fig. \ref{fig:lplm} for the difference of $L_+$ and $L_-$.}
	\label{fig:diskhhe}
\end{figure}

\begin{figure}[H]
	\centering
	\includegraphics[width=0.5\linewidth]{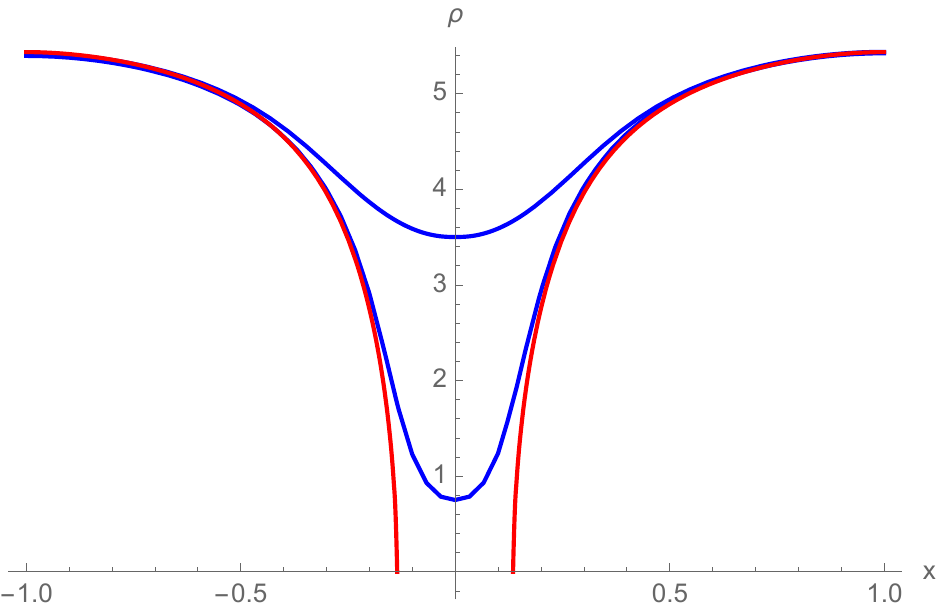}
	\caption{The extremal surfaces with disk radius $L_+\simeq L_-\approx 5.43 > L_c$. It shows one disk radius corresponds to several extremal surfaces for $L>L_c$. We choose the 
one with the minimal area as the RT surface.}
	\label{fig:disksfig}
\end{figure}

Now let us study the HEE of disks for the planar wormhole obtained in sect. 3.2. Because the discussions are similar to those of toy model, we list only key results below. Similar to the toy model, the connected extremal surface exists only if the disk radius is larger than one critical value, $L\ge L_c$. As shown in Fig. \ref{fig:diskhhe}, we have $L_c\approx 4.68$. Besides, As shown in Fig. \ref{fig:disksfig}, one disk radius corresponds to sever extremal surfaces when $L>L_c$. Unlike the toy model with $L_c<L_p$, the area of connected extremal surface, once exists, is always smaller than that of disconnected extremal surface.  See Fig. \ref{fig:diskhhe}, where we have subtracted a universal UV divergence for all areas, i.e., $\frac{L}{ \epsilon} $($\epsilon=10^{-4}$).  Thus, the critical radius $L_c$ is also the phase-transition point $L_p$ of disconnected and connected phases. 

\begin{figure}[H]
	\centering
	\includegraphics[width=0.5\linewidth]{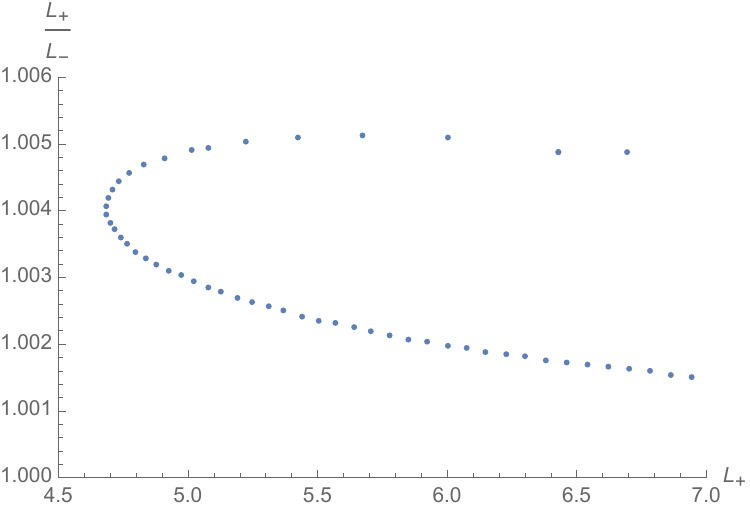}
	\caption{Difference of disk radii in the connected phase. It shows the difference between $L_-$ and $L_+$ is tiny.}
	\label{fig:lplm}
\end{figure}

It is important to note that, due to the asymmetric wormhole solution, the BC (\ref{bdyDC2}) results in slightly different disk radii on either side of the wormhole in the connected phase. See Fig. \ref{fig:lplm} for example, where the difference of $L_+$ and $L_-$ is slight. As illustrated in Fig. \ref{fig:diskhhe}, when $L_+ < L_c \approx 4.68$, only the disconnected phase exists, allowing us to choose $L_+ = L_-$ freely. However, for $L_+ > L_c \approx 4.68$, the connected phase predominates, and $L_-$ is determined by $L_+$, which generally differs from $L_+$. Fortunately, this difference is negligible, as demonstrated in Fig. \ref{fig:lplm}. Similar to the toy model discussed in section 5.1, the connected extremal surface appears only when the disk radius exceeds a critical value, $L > L_c$. Unlike the toy model, the critical disk radius represents the phase-transition point between the disconnected and connected phases, meaning $L_c = L_p$.

\section{Conclusions and Discussions}

The traversable wormholes have recently been obtained in the Einstein-Dirac-Maxwell model without exotic matter. This paper generalizes the discussions to the AdS spacetime and obtains traversable wormholes with spherical and planar topologies. Additionally, we investigate the holographic entanglement entropy of strips and disks for the traversable AdS wormhole. As the size of the strip or disk increases, the Ryu-Takayanagi (RT) surface for entanglement entropy experiences a phase transition, shifting from a disconnected phase to a connected phase. Interestingly, for the disk, the connected extremal surface only exists when the disk radius exceeds a critical value. We verify this novel phenomenon using a toy model, confirming that it also appears in the case of the planar wormhole. Furthermore, we express a desire to explore the hyperbolic wormhole and examine the quantum entanglement of dual conformal field theories (CFTs) in future research.

\section*{Acknowledgements}

We thank Y. Q. Wang and S. W. Wei for valuable comments and discussions. This work is supported by the National Natural Science Foundation of China (No.12275366).

\section*{Appendix A}
This appendix lists the EOMs of Einstein-Dirac-Maxwell model for the AdS wormhole with a spherical topology.

\noindent\hrulefill

\begin{align}\label{gwe1}
	F'(x) &= -\frac{3 F(x) r'(x)}{4 r(x)} + \frac{F(x) r(x) V'(x)^2}{4 N(x)^2 r'(x)} + \frac{4 (F(x)^2 + G(x)^2) (\omega + q V(x)) r(x)}{B(x)^2 N(x)} \nonumber \\
	&\quad + \frac{l^2 F(x) r'(x)}{B(x) N(x) r(x)} - \frac{l^2 G(x) (\omega - m N(x) + q V(x)) r'(x)}{B(x) N(x)} \nonumber \\
	&\quad + \frac{F(x) \left(l^2 - 32 F(x) G(x) l^2 r(x) + \left(16 m (F(x)^2 - G(x)^2) l^2 + 3\right) r(x)^2 \right) r'(x)}{4 l^2 B(x)^2 N(x) r(x)} \tag{A.1}, \\
	G'(x) &= -\frac{3 G(x) r'(x)}{4 r(x)} + \frac{G(x) r(x) V'(x)^2}{4 N(x)^2 r'(x)} + \frac{4 (F(x)^2 + G(x)^2) (\omega + q V(x)) r(x)}{B(x)^2 N(x)} \nonumber \\
	&\quad - \frac{l^2 G(x) r'(x)}{B(x) N(x) r(x)} + \frac{l^2 F(x) (\omega + m N(x) + q V(x)) r'(x)}{B(x) N(x)} \nonumber \\
	&\quad + \frac{G(x) \left(l^2 - 32 F(x) G(x) l^2 r(x) + \left(16 m (F(x)^2 - G(x)^2) l^2 + 3\right) r(x)^2 \right) r'(x)}{4 l^2 B(x)^2 N(x) r(x)} \tag{A.2}, \\
	V''(x) &= \frac{8 q (F(x)^2 + G(x)^2) N(x) r'(x)}{B(x)^2} + \frac{V'(x) \left(r''(x) - \frac{2 r'(x)^2}{r(x)}\right)}{r'(x)} \nonumber \\
	&\quad + \frac{8 \left(-2 F(x) G(x) N(x) + G(x)^2 r(x) \left(2 \omega - m N(x) + 2 q V(x)\right)\right) V'(x) r'(x)}{B(x)^2 N(x)} \nonumber \\
	&\quad + \frac{8 F(x)^2 r(x) \left(2 \omega + m N(x) + 2 q V(x)\right) V'(x) r'(x)}{B(x)^2 N(x)} \tag{A.3}, \\
	N'(x) &= -\frac{r(x) V'(x)^2}{2 N(x) r'(x)} - \frac{N(x) r'(x)}{2 r(x)} + \frac{8 (F(x)^2 + G(x)^2) r(x) (\omega + q V(x)) r'(x)}{B(x)^2} \nonumber \\
	&\quad + \frac{N(x) \left(l^2 - 32 F(x) G(x) l^2 r(x) + \left(16 m (F(x)^2 - G(x)^2) l^2 + 3\right) r(x)^2\right) r'(x)}{2 l^2 B(x)^2 r(x)} \tag{A.4}, \\
	\label{gwe3} B'(x) &= -\frac{B(x) r(x) V'(x)^2}{2 N(x)^2 r'(x)} - \frac{8 (F(x)^2 + G(x)^2) r(x) (\omega + q V(x)) r'(x)}{B(x) N(x)} \nonumber \\
	&\quad + \frac{(l^2 - B(x)^2 l^2 + 3 r(x)^2) r'(x)}{2 l^2 B(x) r(x)} \tag{A.5}.
\end{align}
\noindent\hrulefill

\section*{Appendix B}
This appendix lists the EOMs of Einstein-Dirac-Maxwell model for the planar AdS wormhole. (Here we set $l=1$)

\noindent\hrulefill
\begin{align}\label{pEOMs}
	F'(x) &= \frac{F(x) r(x) V'(x)^2}{4 N(x)^2 r'(x)}
	-\frac{\left(B(x) G(x) - 4 F(x) \left(F(x)^2 + G(x)^2\right) r(x)\right) (\omega - q V(x)) r'(x)}{N(x) B(x)^2} \nonumber \\
	&\quad - \frac{\left(3 F(x) B(x)^2 + 4 m G(x) r(x) B(x) + F(x) \left(16 m F(x)^2 - 16 m G(x)^2 + 3\right) r(x)^2\right) r'(x)}{4 B(x)^2 r(x)}\tag{B.1}, \\
	G'(x) &= \frac{G(x) r(x) V'(x)^2}{4 N(x)^2 r'(x)}
	+\frac{\left(B(x) F(x) + 4 G(x) \left(F(x)^2 + G(x)^2\right) r(x)\right) (\omega - q V(x)) r'(x)}{N(x) B(x)^2} \nonumber \\
	&\quad - \frac{\left(3 G(x) B(x)^2 + 4 m F(x) r(x) B(x) + G(x) \left(16 m F(x)^2 - 16 m G(x)^2 + 3\right) r(x)^2\right) r'(x)}{4 B(x)^2 r(x)}\tag{B.2}, \\
	V''(x) &= \frac{8 q N(x) \left(F(x)^2 + G(x)^2\right) r'(x)^3 - 8 \left(F(x)^2 + G(x)^2\right) r(x) (\omega - 2 q V(x)) V'(x) r'(x)^2}{N(x) B(x)^2} \nonumber \\
	&\quad + \frac{V'(x) \left(\left(r(x) r''(x) - 2 r'(x)^2\right) B(x)^2 + 8 m \left(F(x)^2 - G(x)^2\right) r(x)^2 r'(x)^2\right)}{r(x) B(x)^2 r'(x)}\tag{B.3}, \\
	N'(x) &= -\frac{r(x) V'(x)^2}{2 N(x) r'(x)}
	-\frac{8 \left(F(x)^2 + G(x)^2\right) r(x) (\omega - q V(x)) r'(x)}{B(x)^2} \nonumber \\
	&\quad - \frac{N(x) \left(B(x)^2 + \left(-16 m F(x)^2 + 16 m G(x)^2 - 3\right) r(x)^2\right) r'(x)}{2 B(x)^2 r(x)}\tag{B.4}, \\
	B'(x) &= \frac{r(x) \left(3 N(x) - 16 q \left(F(x)^2 + G(x)^2\right) V(x)\right) r'(x)}{2 N(x) B(x)}
	+ B(x) \left( -\frac{r(x) V'(x)^2}{2 N(x)^2 r'(x)} - \frac{r'(x)}{2 r(x)} \right).\tag{B.5}
\end{align}



\end{document}